\documentclass[10pt,twocolumn,superscriptaddress,aps,prl,showpacs,preprintnumbers,amsmath,amssymb,floatfix]{revtex4}

\usepackage[latin9]{inputenc}
\usepackage{color}

\usepackage{graphicx}
\usepackage{esint}

\newcommand{\ket}[1]{| {#1} \rangle}
\newcommand{\expect}[1]{\langle {#1} \rangle}

\makeatletter
\@ifundefined{textcolor}{}
{%
 \definecolor{BLACK}{gray}{0}
 \definecolor{WHITE}{gray}{1}
 \definecolor{RED}{rgb}{1,0,0}
 \definecolor{GREEN}{rgb}{0,1,0}
 \definecolor{BLUE}{rgb}{0,0,1}
 \definecolor{CYAN}{cmyk}{1,0,0,0}
 \definecolor{MAGENTA}{cmyk}{0,1,0,0}
 \definecolor{YELLOW}{cmyk}{0,0,1,0}
}

\usepackage{graphicx}
\usepackage[english]{babel}
\usepackage{amsmath}
\usepackage{amssymb}
\usepackage{bm}

\newcommand{\sgn}{{\mbox{ sgn}}}

\begin{document}
\title{Parafermionic zero modes in ultracold bosonic systems}
\author{M. F. Maghrebi}
\email[Corresponding author: ]{magrebi@umd.edu}
\affiliation{Joint Quantum Institute, NIST/University of Maryland, College Park, Maryland 20742, USA}
\affiliation{Joint Center for Quantum Information and Computer Science, NIST/University of Maryland, College Park, Maryland 20742, USA}
\author{S. Ganeshan}
\affiliation{Joint Quantum Institute, NIST/University of Maryland, College Park, Maryland 20742, USA}
\affiliation{Condensed Matter Theory Center and Physics Department, University of Maryland, College Park, Maryland 20742, USA}
\author{D. J. Clarke}
\affiliation{Condensed Matter Theory Center and Physics Department, University of Maryland, College Park, Maryland 20742, USA}
\author{A. V. Gorshkov}
\affiliation{Joint Quantum Institute, NIST/University of Maryland, College Park, Maryland 20742, USA}
\affiliation{Joint Center for Quantum Information and Computer Science, NIST/University of Maryland, College Park, Maryland 20742, USA}
\author{J. D. Sau}
\affiliation{Joint Quantum Institute, NIST/University of Maryland, College Park, Maryland 20742, USA}
\affiliation{Condensed Matter Theory Center and Physics Department, University of Maryland, College Park, Maryland 20742, USA}

\begin{abstract}
 Exotic topologically protected zero modes with parafermionic statistics (also called fractionalized Majorana modes) have been proposed to emerge in devices fabricated from a fractional quantum Hall system and a superconductor. The fractionalized statistics of these modes takes them an important step beyond the simplest non-Abelian anyons, Majorana fermions. Building on recent advances towards the realization of fractional quantum Hall states of bosonic ultracold atoms, we propose a realization of parafermions in a system consisting of Bose-Einstein-condensate trenches within a bosonic fractional quantum Hall state. We show that parafermionic zero modes emerge at the endpoints of the trenches and give rise to a topologically protected degeneracy.
We also discuss methods for preparing and detecting these modes.
\end{abstract}


\maketitle

In recent years the concept of topological order has revolutionized the way we understand quantum phases of matter. Topological phases in one- and two-dimensional systems are particularly interesting as the nontrivial exchange statistics of particles allows for exotic states of matter. For example, Majorana zero modes can emerge at boundaries of one-dimensional topological superconductors  \cite{Kitaev01} or in two-dimensional semiconductor heterostructures \cite{Sau10,Lutchyn10,Alicea10}; see also suggestive experimental signatures in Refs.~\cite{Mourik12,das12,rokhinson12,deng12,finck13,churchill13}.
These topological modes have been a subject of intense interest due to their potential applications in quantum computation, although they do not support universal quantum computing with braiding alone
\cite{freedman98,Freedman02,kitaev03,freedman03,parsa08,nayak08}. In
two dimensions, certain fractional quantum Hall (FQH) states have been proposed to manifest emergent non-Abelian excitations with universal braiding statistics \cite{read99,fendley09,bishara08}.
However, an experimental confirmation of such emergent non-Abelian anyons has so far remained elusive \cite{pan99,xia04,pan08}.

Recently, it has been proposed that one can engineer non-Abelian excitations by adding defects in the form of ferromagnet-superconductor interfaces at the edges of adjacent Abelian FQH states~\cite{Clarke13,Lindner12,Cheng12,Vaezi13}. The domain wall at their interface binds exotic zero modes with parafermionic commutation relations. These parafermionic zero modes are associated with a ground-state  degeneracy that is exponential in the number of domain walls (defects). As with any non-Abelian anyon, the exchange of the defects binding these parafermionic operators generates a unitary rotation in the ground-state subspace. However, the set of operations for quantum computation available through such exchanges is richer than that available through Majorana exchange \cite{Clarke13}. Parafermions may also be realized in bilayer quantum hall systems, where the role of the superconducting- (or BEC-) induced coupling is played by an interlayer tunneling term \cite{barkeshli13b,Barkeshli14d}. For a recent proposal on parafermions, see also Ref.~\cite{Meng15}.

While existing proposals are based on an experimentally challenging combination of FQH and superconducting systems of electrons, rapid advances towards creating a bosonic FQH state open new opportunities to realize topologically nontrivial states in the context of ultracold bosonic systems \cite{Ruseckas05,Lin09,Gerbier10,Bloch12}.
Unlike fermionic systems, where a condensed state requires pairing of fermions, systems of ultra-cold bosons
form Bose-Einstein condensates (BECs) without any additional pairing interaction. Thus, if a FQH state is realized in bosons, adding a Bose \ condensed state to such a system can be expected to be simpler than in the corresponding fermionic implementation.

\begin{figure}[t]
  \begin{center}
  \includegraphics[width=8.5cm]{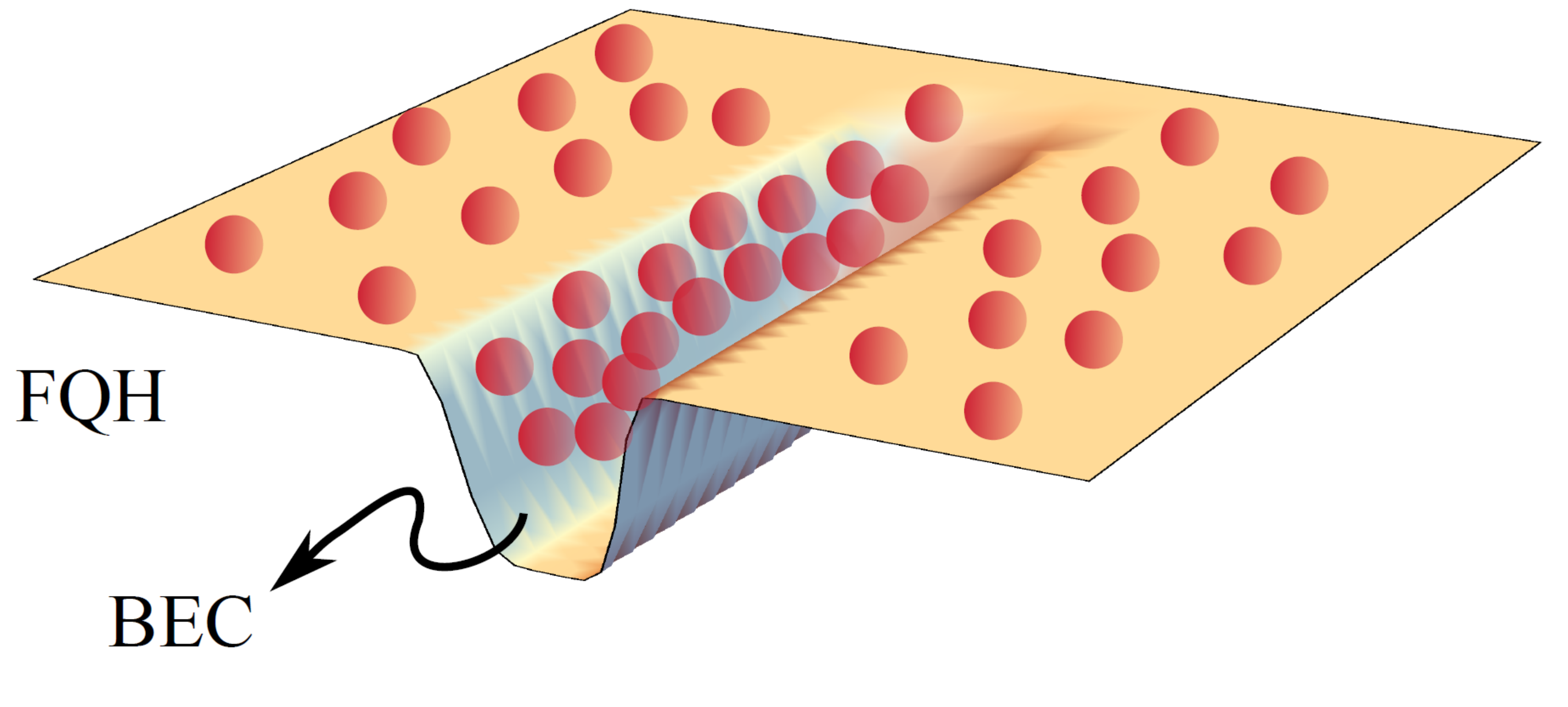}
  \caption{A quasi-one-dimensional finite trench, i.e.\ a potential dip, is created by spatially modulating the intensity of the laser used to create the dipole trapping potential confining the atoms to the plane (a cut midway along one trench is shown).
Within the trench, the two-dimensional density of trapped bosons deviates from the FQH filling fraction and gives rise to a BEC.}
  \label{Fig. BECTrench}
  \end{center}
\end{figure}

In this Letter,
we propose to realize a bosonic analog of a FQH-superconductor system by inserting a trench containing a BEC in the middle of a bosonic FQH system,
as shown in Fig.~\ref{Fig. BECTrench}. Such a trench of BEC can be created by introducing a potential well that would trap a high density of bosons as compared to the FQH region outside. We will show that the introduction of such a BEC
trench induces a novel state in the quantum Hall edge with a pairing gap similar to the states with superconductors
in contact with the FQH  system. Further we will show that systems containing a pair of such trenches feature a ground state degeneracy arising from
the topology of the underlying quantum Hall system.
We explicitly construct fractionalized Majorana fermions
as localized zero modes at the endpoints of the trenches.

\emph{Model.}---Let us consider a bosonic FQH state with a filling fraction $\nu=1/m$, where $m$ is an even integer. While the bulk of the FQH fluid is incompressible, its boundaries
support gapless edge excitations with fractionalized charge, or boson number in this case, and statistics \cite{Tao83,Arovas84}. The edge modes of a FQH state can be described by a bosonized field $e^{i\varphi(\xi)}$ that carries boson number $1/m$ and satisfies a nontrivial commutation relation \cite{Wen04}
\begin{equation}
  e^{i\varphi(\xi)} e^{i\varphi(\xi')}=e^{i\frac{\pi}{m} {\footnotesize \sgn}(\xi-\xi')}e^{i\varphi(\xi')} e^{i\varphi(\xi)},
\end{equation}
where $\xi$ and $\xi'$ are the coordinates along the boundary. Equivalently, the chiral field $\varphi$ satisfies the chiral commutation relations.
The creation operator for bosons is given by $e^{i m\varphi(\xi)}$; note that bosonic operators commute according to the above algebra. Furthermore, the density of
bosons on the edge is given by $\rho(\xi)=\partial_\xi \varphi/2\pi$. The Hamiltonian $H_0=\frac{m v}{4\pi} \int_\Gamma   d\xi \,\, (\partial_\xi \varphi)^2$, along with the chiral commutation $[\varphi(\xi), \partial_{\xi'}\varphi(\xi')]=\frac{2\pi i}{m}\delta(\xi-\xi')$ describes a free chiral edge mode propagating at velocity $v$ along the boundary $\Gamma$ of the FQH  state.

As shown in Fig.\ \ref{Fig. BECTrench}, we consider a BEC trench  inside a bosonic FQH state that can be introduced by varying
the potential. The topological phase realized by this FQH system will be characterize by a topological degeneracy. To
create a Hilbert space that allows us to access this degeneracy, we will need to consider a system containing two
such trenches in the FQH system [two copies of the trench in Fig.~\ref{Fig. BECTrench}].
 The total Hamiltonian describing the boundaries of the FQH state at the edges of the two BEC trenches is given by
\begin{align}
H=H_1(\varphi_1, Q_1)+H_2(\varphi_2, Q_2)\label{model}\\
\text{with}\ \  Q_1+Q_2=0 \mod 1,\nonumber
\end{align}
where the subscripts denote the trenches,
and $Q_i$ is the fractional number (``charge") of quasiparticles modulo 1
on the edge of the $i$th trench. While each edge can have a fractional charge, the total  number of quasiparticles must add up to an integer bosonic charge, i.e.\ $Q_1+Q_2=0$ mod 1.
We show that, under certain conditions, the Hilbert space of this system manifests degenerate ground states characterized by parafermionic zero modes at the ends of the BEC trenches. The defects at each end of the trench thus act as non-Abelian anyons.  This is the central result of this Letter. We also analyze the robustness of this degeneracy to realistic experimental imperfections.

\emph{Single-trench Hamiltonian.}---For simplicity, we first focus on a single edge, with the Hamiltonian $H_1(\varphi_1,Q_1)$, without the constraint in Eq.~(\ref{model}), but will impose it later in our discussion of the
degeneracy of the two-trench system. For now, we also drop the subscript 1 for notational simplicity.
The tunneling between the BEC field $\Psi$ and the edge states on the trench is described by
\begin{equation}\label{Eq. FQH coupled to BEC}
    H_{\rm tun}=-\Delta \int_{\Gamma} \!\!d\xi \,\, e^{im \varphi(\xi)} \Psi(\xi) +{\rm h. c. }
\end{equation}
Note that only a boson, i.e.\ $m$ quasiparticles bound together, can directly tunnel to the BEC. The effect of the
BEC on the edge can be understood by first expanding the BEC field as $\Psi(x)=\Psi_0+\delta\Psi(x)$, where $\delta\Psi(x)$
is the boson fluctuation field in a three-dimensional BEC, and is described by a quadratic Hamiltonian.
The fluctuations of $\delta\Psi(x)$ can be integrated out to obtain an effective self-energy for the edge induced by
the BEC, $\Sigma=\Sigma^{(1)}+\Sigma^{(2)}$, where $\Sigma^{(1)}=-\Psi_0 \Delta\int_{\Gamma} \!d\xi \,\, e^{im \varphi(\xi)} +h.c.$ is linear in $\Delta$, while $\Sigma^{(2)}$ is second order in $\Delta$ and is proportional to
the Green's function of the BEC as we shall discuss below.

The lowest order self-energy term $\Sigma^{(1)}$ does not have the form of a cosine term in the sine-Gordon Hamiltonian \cite{Giamarchi03}
known to qualitatively modify the chiral edge state by opening a gap.
The effect of this unconventional term can be eliminated by shifting the density $\rho(\xi)$  of the edge via $\rho(\xi)\rightarrow \rho(\xi)+\rho_0$
for some nonzero $\rho_0$. The density of particles at the edge can be tuned by changing the trap potential,
which in turn changes the chemical potential in the BEC. Such a shift can be accommodated by shifting the chiral
boson field as $\varphi(\xi)\rightarrow\varphi(\xi)+  2 \pi \rho_0 \xi$.
This shift introduces a spatial dependence into the lowest order self-energy $\Sigma^{(1)}\rightarrow
-\Psi_0 \Delta\int_{\Gamma} \!\!d\xi \,\, e^{i 2 \pi m \rho_0 \xi} e^{im \varphi(\xi)} +h.c.$, and leads to a momentum mismatch \cite{kanemomentum}, which suppresses the effect of the first-order self-energy $\Sigma^{(1)}$.

Therefore, at a nonzero
$\rho_0$, the effect of the self-energy is dominated by the
second-order term $\Sigma^{(2)}$, which can be written as $\Sigma^{(2)}=-\frac{\Delta^2}{2} \!\! \int_{\Gamma} \!d\xi d\xi' e^{\pm i m ( \varphi(\xi)+ 2 \pi \rho_0 \xi) \mp i m( \varphi(\xi')+ 2 \pi \rho_0 \xi')} \times \expect{\delta\Psi_\alpha(\xi)\delta\Psi_\beta^\dagger(\xi')}$,
where $\alpha,\beta=0,1$ corresponds to the
upper/lower sign of the corresponding $\varphi$,
and $(\delta\Psi_0\quad \delta\Psi_1)=(\delta\Psi\quad\delta\Psi^\dagger)$. The correlator in $\Sigma^{(2)}$ is related to the Green's function $G_{\alpha\beta}\left(\bm r(\xi),\bm r(\xi')\right)$ of the BEC, where $\bm r (\xi)$ is the position vector of the point $\xi$. A more careful discussion of the perturbation theory is given in the Supplemental Material \cite{supp}. The Green's functions are highly peaked at $\bm r(\xi) \approx \bm r(\xi') $ and can hence be split into a dominant local part and a perturbatively small non-local part.
In the remainder of the paper, we will first discuss the physics of the dominant local part where $|\bm r(\xi)-\bm r(\xi')|\lesssim W$ with $W$ the width of the trench, and comment on the effect
of the perturbative non-local part near the end.

\begin{figure}[t]
  \centering
   \includegraphics[width=8cm]{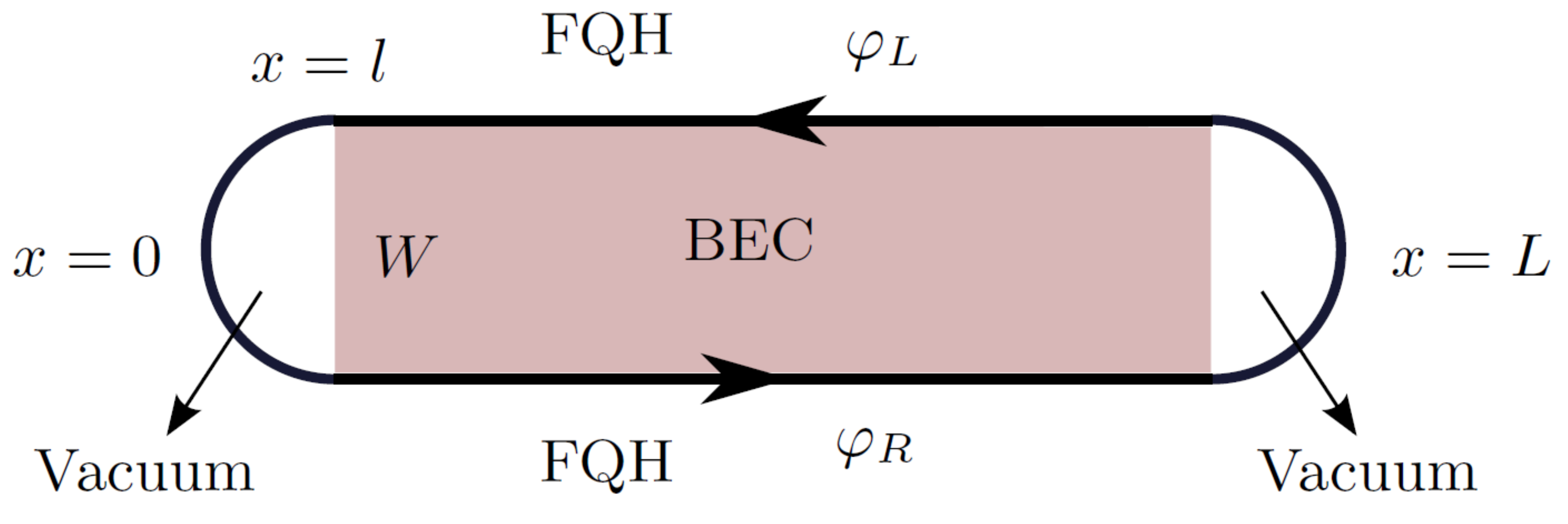}
  \caption{The top-view of a BEC trench within a FQH state. The trapping potential is engineered such that the boson density is uniform across the shaded area but  vanishes in a small region of size $l$ near the endpoints. Quantum fluctuations of the BEC couple opposite edges in the shaded region.}
  \label{Fig. BECTrench cartoon}
\end{figure}

As shown in Fig.\ \ref{Fig. BECTrench cartoon},
pairs of points directly across the trench interact through $\Sigma^{(2)}$ since they are separated by a distance comparable to $W$. It is
convenient to define the coordinate $x\in [0,L]$ along
the trench of length $L$, together with the left- and right-moving fields
$\varphi_{L}(x)=\varphi(x)$ and $\varphi_R(x)=\varphi(2L-x)$
propagating along the top and bottom edge, respectively. With this definition, the oscillating phase in $\Sigma^{(2)}$
has a straightforward interpretation: For example, taking $\xi = x$ and $\xi'=2L-x'$,
the phase corresponding to the Green's function $G_{01}$ becomes $e^{i\rho_0(x-x')}$.
The presence of the latter term ensures momentum conservation \cite{kanemomentum}; only such combinations
contribute significantly to $\Sigma^{(2)}$. Physically the only terms that satisfy these constraints are normal particle-particle scattering terms
on the same edge or anomalous particle-hole scattering between opposite edges. The former simply renormalizes the velocity, while the
latter is analogous to superconducting pairing.
Taking the above constraints into account, the dominant, and local, contribution to the effective interaction is
\begin{align}\label{Eq. Eff Hamiltonian}
V_{\rm eff}=- \lambda\int_l^{L-l} \!\!dx \, \cos(2m\phi),
\end{align}
where we have defined the (non-chiral) fields $\phi(x)$ and $\theta(x)$ by $\varphi_{L/R}(x)=\phi(x)\mp\theta(x)$, and the coupling coefficient $ \lambda \propto \Delta^2$. The new non-chiral fields are self-commuting at all points, but satisfy
\mbox{$[\phi(x),\theta(x')]=i(\pi/m)\Theta(x-x')$.}

The term $\cos(2m\phi)$ represents the effective pairing between the edges and is reminiscent of the sine-Gordon model \cite{Giamarchi03}.
The resulting  Hamiltonian that describes the BEC trench (Fig.~\ref{Fig. BECTrench cartoon})     can be written as
 \begin{equation}\label{Eq: total Hamiltonian}
     H_{\rm eff}\approx\frac{mv}{2\pi} \int_{0}^{L} \!\!\! dx\!\left[ (\partial_x\phi)^2 + (\partial_x\theta)^2\right]-\lambda \int_l^{L-l}\!\!\!\!dx \cos (2m\phi),
 \end{equation}
where the continuity of
$\partial_x\varphi(x)$ near $x\sim 0,L$ transforms into the boundary condition $\partial_x \phi(x=0,L)=0$.
We note that, in addition, we are required to preserve the boundary conditions for $\theta$ at the endpoints: $\theta(x=L)=0$ follows from the definition, while $\theta(x=0)=\pi N/m$, where $N$ is the total number of quasiparticles. The latter is due to the fact that the total density of
bosons on both edges is $\rho= -\partial_x \theta/\pi$.
However, similar to other restrictions on the Hilbert space, we will restore these boundary conditions at the end of the calculation.
At large $\lambda>0$, the sine-Gordon model [Eq.\ (\ref{Eq: total Hamiltonian})] supports several
ground states $\ket{p}$ characterized by the expectation values
 \begin{equation}\label{Eq. phi-n values}
   \expect{p|e^{i\phi(x)}|p} \approx e^{i\pi p/m},  \qquad p=0,1, \cdots, 2m-1,
 \end{equation}
for $x$ away from the edges and in the limit of large $L$ \cite{Giamarchi03}.

To restore the appropriate boundary conditions for $\theta(x=0,L)$, we notice that $\theta(x=L)$ commutes with $\phi(x)$ and
$H_{eff}$, hence it can be set to any value [$\theta(x=L)=0$ in our case] without consequence for the single trench. On the other hand, $\theta(x=0)=\pi N/m$
obeys a non-trivial commutation relation with $\phi(x)$ such that $[N,\phi]=-i$; however, $R=e^{i\frac{\pi}{m}N}$ commutes with $H_{eff}$.
The operator $R$, despite being a symmetry of the Hamiltonian, transforms $\phi(x)$ in a non-trivial way as
\begin{equation}
    R \phi(x)R^\dagger =\phi(x)+\frac{\pi}{m}\,.
\end{equation}
It then follows that, including the boundary conditions, different values of $\phi$ in Eq.~(\ref{Eq. phi-n values}) indeed correspond to the same energy.
Exploiting the above symmetry, we can describe the ground states in a basis that also makes the operator $R$ diagonal as
\begin{equation}
|n\rangle =\sum_{p=0}^{2m-1} e^{i \pi n p/m} \left | p\right\rangle,
\end{equation}
which satisfy $R\ket{n}=e^{i\pi n/m}\ket{n}$. Physically, $n$ corresponds to the number of quasiparticles modulo pairs of
bosons on the edge of a given trench. In our model thus far, this is a well-defined quantum number since the cosine term in Eq.\ (\ref{Eq: total Hamiltonian})
transfers only pairs of bosons from the condensate. Note that $N$ takes $2m$ distinct eigenvalues. This allows us to associate a degeneracy of $\sqrt{2m}$ with each of the endpoints of the BEC trench (the `defects') in the limit of a large number of trenches where restrictions on the Hilbert space may be safely ignored. We will discuss Hilbert space constraints for a small number of trenches in the next section.
Finally we remark that the fractional part of the boson number on a trench, which is invariant under the addition of single bosons, is $Q=N/m$ mod 1, while
$Q_b\equiv [N/m-Q]\in\{0,1$\} defines the boson parity.
The $\sqrt{2m}$
degeneracy thus also requires protection by
a $Z_2$ symmetry due to boson parity.

\emph{Degeneracy.}---So far we have focused on the spectrum of a single trench without the physical constraints of the Hilbert space. With a finite number of trenches, the total fractional charge $Q$ should be $0 \mod 1$
\footnote{This constraint can be relaxed by allowing fractional excitations on the edge of the quantum Hall regime, but this will lead to a degeneracy that decays polynomially in the system size.}.
Considering the double trench model of Eq.~(\ref{model}), the fractional boson numbers on the two trenches satisfy $Q_1=- Q_2 = 0,1/m,\cdots, 1-1/m$, while their boson parities $Q_{b,1}=0,1$ and $Q_{b,2}=0,1$ are unconstrained, which yields a total degeneracy of $D=4m$ for two trenches. More generally, a system of $k$ trenches with boson-parity conservation intact has the degeneracy
\begin{equation}\label{Eq. degeneracy}
D=2(2m)^{k-1}.
\end{equation}
  We shall discuss the (topological) robustness of this degeneracy, which partially survives the $Z_2$ symmetry breaking, after we explicitly construct operators that span the $4m$ degeneracy present in the two trench case.

\emph{Parafermion operators.}---We now construct the operators spanning the above two-trench degeneracy. As remarked earlier, this degeneracy is spanned by the exchange of quasiparticles between the endpoints of the trenches. Let  $U^\dagger_{l}$ ($U^\dagger_r$) be the exchange operator between the left (right) endpoints of the two trenches. They can be expressed as
 \begin{equation}\label{Eq. exch op}
 U^\dagger_{l,r}\equiv T_{l,r}^\dagger B_{l,r},
 \end{equation}
 where $T^\dagger$ and $B^\dagger$ represent quasiparticle vertex operators acting on the top trench and bottom trench, respectively,  and projected onto the ground-state sector. For example, $T^\dagger_{l,r} = e^{i\varphi_1(x=0,L)}$ adds a quasiparticle on one of the two ends of the first trench with the projection onto the ground state implicit. The operator $B$ on the second trench can be defined similarly by $\varphi_1 \to \varphi_2$; however, we must insure that the exchange of quasiparticles between the two trenches respects the exchange statistics. This can be done consistently by including a Klein factor as $B^\dagger_{l,r} = e^{i\pi N_1/m} e^{i\varphi_2(x=0,L)}$, where $N_1$ is the number of quasiparticles on the first trench.
 With this construction, we now focus on the operator $T$ defined above. Note that the chiral field at the ends is given by $\varphi_1= \phi_1-\theta_1$. As we discussed earlier, $\theta_1(L)=0$ and $\theta_1(0)= \pi N_{\theta,1}/m$ mod $2\pi$ with $N_{\theta,1} \equiv m(Q_{b,1}+Q_1)$ the total number of quasiparticles modulo $2m$ on the first trench. Since we are interested in the ground state sector, the field $\phi$ is roughly assumed to be pinned according to Eq.~(\ref{Eq. phi-n values}) over the entire edge as $\phi_1(0)\approx\phi_1(L)\approx 2\pi N_{\phi,1}/{m}$ with integer $N_{\phi,1}$. We find
\begin{align}\label{Eq. Zero modes}
T_l^\dagger=e^{i \frac{\pi}{m} (N_{\phi,1}-N_{\theta,1})}, \qquad
T_r^\dagger=e^{i \frac{\pi}{m} N_{\phi,1}},
\end{align}
which satisfy the algebra
\begin{equation}\label{Eq. parafermion ops}
  (T_{l,r}^\dagger)^{2m}=1, \qquad T_l^\dagger T_r^\dagger = e^{-i\frac{\pi}{m} }T_r^\dagger T_l^\dagger.
\end{equation}
These relations describe parafermions, a generalization of the fermionic algebra \cite{Fendley12,Clarke13,Lindner12,Cheng12,Vaezi13}, see also Ref.~\cite{Nussinov12}.

The operators $B_{l,r}$ can be defined similar to Eq.~(\ref{Eq. Zero modes}) by including the Klein factor $\exp({i\pi N_{\theta,1}/m})$ explained above, and with $N_{\theta,1} \to N_{\theta,2}=m(-Q_1+Q_{b,1})$ and $N_{\phi,1}\to N_{\phi,2}$. The parafermionic algebra implies that $(U^\dagger)^{2m}=1$, which yields $2m $ degenerate ground states in a sector with a fixed total parity [the operator $U$ does not change the total parity according to Eq.~(\ref{Eq. exch op})]. With the two-fold degeneracy due to the total parity, one recovers the full $4m$ degeneracy of the system.

\emph{Robustness.}---Heretofore, we have focused on the Hamiltonian in Eq.~(\ref{Eq: total Hamiltonian}). In principle, however, a single boson can tunnel to or from the BEC, which might arise from terms of the type $V = \cos [m \phi(x=0,L)]$ near the endpoints of the trench.
In fact, introducing the perturbation $V$ within first order degenerate perturbation theory reduces the $4m$ fold degeneracy to $D'=m/2$; see the Supplemental Material \cite{supp} for details.
While such a term clearly breaks the boson parity symmetry (hence $(Q_{b,1}, Q_{b,2})$ are no longer good quantum numbers), it also breaks a two-fold degeneracy of the fractional number of quasiparticles $Q_1$.
The remaining $m/2$-fold degeneracy is topologically protected, and we find in general that the degeneracy for $k$ trenches is given by
\begin{equation}\label{Eq. quantum d}
D'=(m/2)^{k-1}.
\end{equation}
This gives a quantum dimension of $\sqrt{m/2}$ for the defects ending the BEC trenches.
However, if the vacuum regions (Fig.~\ref{Fig. BECTrench cartoon}) are sufficiently large, single-boson tunneling is suppressed, and one recovers the degeneracy in Eq.~(\ref{Eq. degeneracy}).

We also briefly remark that the long-range fluctuations of the BEC can be considered effectively as long-range tunneling, and can provide another mechanism to break the boson-parity symmetry, while the quantum dimension in Eq.~(\ref{Eq. quantum d}) will not be affected.

\emph{Preparation and detection.}---There is some evidence that a $\nu=1/2$ fractional Chern insulator has a continuous transition to a BEC, which allows for quasi-adiabatic preparation of the former \cite{barkeshli14b}. A similar procedure may exist for FQH states. The parafermions in the FQH state can be prepared by starting with a small island which can be grown to a trench in linear time by modulating the laser beams. Furthermore, Bragg spectroscopy can provide direct information about the topological phase of the system. For example, one should observe a zero bias peak at the endpoints of the trench \cite{yao13c}.
They can also be probed using braiding, which corresponds to the topologically protected manipulation of the underlying quantum information and which requires dynamically changing the geometry of the system, bringing different sets of parafermionic edge modes in close proximity to each other \cite{Lindner12,Clarke13}. Such dynamical changes can easily be achieved by dynamically changing the laser beam used to create the BEC trenches.

It is worth pointing out that a theoretically simpler but experimentally more challenging approach would proceed by analogy with Ref.\ \cite{jiang11c}:
A BEC of diatomic molecules can be coherently dissociated into pairs of atoms, which readily gives the effective Hamiltonian in Eq.\ (\ref{Eq. Eff Hamiltonian}).

\emph{Conclusion.}---In this work, we have considered a BEC trench in a FQH liquid, and showed that, in a certain regime, the combined system is in a topological phase, which is identified by the zero mode operators at the endpoints of the trench. These zero modes are shown to be parafermions, a generalization of the usual fermionic or bosonic algebra. We have also derived the topological degeneracy of the parafermionic modes, and examined their robustness against local perturbations.

\begin{acknowledgments}
We thank J.\ Alicea, M.\ Barkeshli, K.\ Shtengel, R.\ Mong, and N.\ Cooper for discussions. MFM and AVG gratefully acknowledge support by AFOSR, NSF-JQI-PFC, NSF PIF, ARO, ARL, and AFOSR MURI.
SG gratefully acknowledges support by NSF-JQI-PFC and LPS-CMTC.
DJC gratefully acknowledges support by LPS-CMTC and Microsoft.
JDS gratefully acknowledges the University of Maryland, the Condensed Matter Theory Center, and the Joint Quantum Institute for startup support.
\end{acknowledgments}


\newpage

\begin{centering}
{\hskip .6in \Large \bf Supplemental Material}
\end{centering}

\renewcommand{\thesection}{S\arabic{section}}
\renewcommand{\theequation}{S\arabic{equation}}
\renewcommand{\thefigure}{S\arabic{figure}}
\setcounter{equation}{0}
\setcounter{figure}{0}

\section{Effective interaction to the second order}
We start with the interaction in Eq.~(3). It is more convenient to work in Euclidean space where the interaction takes the form
\begin{equation}
  -\Delta \int d\tau  d\xi  e^{im \varphi(\tau,\xi)} \Psi(\tau, \xi)+c.c.
\end{equation}
We expand the condensate field as $\Psi=\Psi_0 +\delta \Psi$ similar to the main text.
The effective interaction, after integrating out the BEC field, is formally given by
\begin{align}
  e^{-\Theta[\varphi(\tau,\xi)]}=&e^{\Psi_0\Delta \int d\tau d\xi e^{im \phi} +c.c.} \nonumber \\
  &\times \left\langle e^{\Delta \int d\tau  d\xi  e^{im \varphi(\tau,\xi)} \delta\Psi(\tau, \xi)  +c.c.}\right\rangle_{\delta\Psi},
\end{align}
where the average $\langle \cdot\rangle_{\delta \Psi}$ is taken over the fluctuations of $\delta \Psi$. To  first order, the effective interaction is given by the prefactor in the above equation. To  second order, the effective interaction is
\begin{align}
  &\Theta^{(2)}[\varphi]=-\frac{\Delta^2}{2} \int d\tau d\tau'd\xi d\xi' \,\, \times \nonumber \\
  \Big[&e^{im (\varphi(\tau, \xi)+ \varphi(\tau',\xi'))}e^{i 2 \pi m \rho_0 (\xi+\xi')}G_{01}[\tau,\bm r,\tau',\bm r']+c.c. \nonumber \\
  &e^{im (\varphi(\tau,\xi)-\varphi(\tau',\xi'))}e^{i 2 \pi m \rho_0  (\xi-\xi')}G_{00}[\tau,\bm r,\tau',\bm r']+c.c.\Big],
\end{align}
where $\bm r=\bm r(\xi)$ and $\bm r'=\bm r(\xi')$, and
\begin{align}
  G_{00}(\tau,\bm r,0,\bm 0)&=\expect{\delta\Psi(\tau, \xi)\delta\Psi^\dagger(0,0)},\nonumber \\
  G_{01}(\tau,\bm r,0,\bm 0)&=\expect{\delta\Psi(\tau, \xi)\delta\Psi(0,0)},
\end{align}
are the normal and anomalous Green's functions of the BEC, respectively. While these functions are nonlocal both in space and (imaginary) time, they are highly peaked around $(\tau,\bm r)\to 0$, and thus the interaction is assumed to be dominantly local plus small non-local terms. In the second-order perturbation theory treatment of the main text, we have neglected the time-dependence of the Green's functions to simplify the presentation. With the interaction approximated to be local in time as well as space, one can write the effective interaction as
\begin{equation}
  \Theta[\varphi(t,x)]=\int d\tau \, \Sigma[\varphi(\tau,x)]=\int d\tau dx \, V_{\rm eff}\left(\varphi(\tau,x)\right),
\end{equation}
where $x$ is the position along the trench defined in the main text,
and $V_{\rm eff}$ is the effective Hamiltonian, which, after dropping the momentum-mismatched terms, becomes Eq. (4) in the main text.

\section{Derivation of the quantum dimension in the absence of boson-parity symmetry}

In this Section, we perform first-order perturbation theory in the absence of boson-parity symmetry  and derive the quantum dimension given in Eq.~(13).

We begin by consider a single trench without any constraints, and only later implement the constraint on the physical Hilbert space. The parafermion operators inroduced in the main text allow us to define a state with $n$ quasiparticles on the edge of the trench as
\begin{equation}
  |n\rangle = \left(T^\dagger_r\right)^n|0\rangle,
\end{equation}
where $|0\rangle$ is the state with no fractional charge. The single-boson operators on the two ends of the trench are defined as
\begin{equation}
  b_{l/r}= \left(T^\dagger_{l/r}\right)^m|0\rangle,
\end{equation}
at the corresponding left/right endpoints. We have
\begin{equation}
  b_r |n\rangle =|n+m\rangle,
\end{equation}
by definition, and
\begin{align}
  b_l |n\rangle&= b_l \left(T^\dagger_r\right)^n |0\rangle =(-1)^n\left(T^\dagger_r\right)^n b_l |0\rangle \nonumber\\
  &=\sigma (-1)^n |n+m\rangle,
\end{align}
where $\sigma= \langle m|b_l |0\rangle =\langle 0|b_r^\dagger b_l |0\rangle$; in deriving the above equation, we used the commutation relation of parafermion operators in Eq.~(12) of the main text, and specifically $b_l T^\dagger_r=-T^\dagger_r b_l$. We also note that the quantum numbers in brackets should be understood modulo $2m$.

Using the definitions given in the main text, one can see that the operators $b_l$ and $b_r$ are Hermitian, and $\sigma=(-1)^{m/2}$ is a real number ($m$ is even). Therefore, a perturabtion of the Hamiltonian that adds a single boson, and thus breaks the boson parity, has the form
\begin{equation}
  H'= \alpha b_l +\beta b_r,
\end{equation}
with $\alpha$ and $\beta$ being arbitrary real numbers. This Hamiltonian mixes the states $|n\rangle$ and $|n+m\rangle$ as
\(
  \langle n+m|H'|n\rangle = (-1)^n\sigma \alpha +\beta.
\)
Thus, the eigen-energies are split as
\[
\pm \left[(-1)^n\sigma \alpha +\beta\right],
\]
which implies that the number of minimum-energy states is reduced by a factor of 4, from $2m$ (see the main text) to $m/2$.

In the next step, we consider two trenches, and impose the physical constraint that the total fractional charge is trivial, i.e., $n^{(1)}+n^{(2)}=0$ mod $m$ where the superscripts denote the two trenches. In particular, this implies that $(-1)^{n^{(1)}}=(-1)^{n^{(2)}}$. A similar perturbative treatment yields the energy splitting
\begin{equation}
  \pm \left[(-1)^{n^{(1)}}\sigma \left(\alpha^{(1)}+\alpha^{(2)}\right) +\left(\beta^{(1)}+\beta^{(2)}\right)\right],
\end{equation}
with $\alpha^{(i)}$ and $\beta^{(i)}$ the corresponding perturbation parameters for each trench.
The splitting of energy reduces the total degeneracy down to $m/2$. (Note that there is no freedom in choosing the relative parity of $n^{(1)}$ and $n^{(2)}$ due to the mixing of $n^{(i)}$ and $n^{(i)}+m$ in the true ground states of the perturbed Hamiltonian.) A straightforward generalization to arbitrary number of trenches shows that the quantum dimension of the endpoint defects in the absence of boson-parity symmetry is given by Eq.~(13) of the main text.

\end{document}